\documentclass{appolb}
\usepackage{graphicx}

\begin{document}
\title{First tests of the full SIDDHARTA-2 experimental apparatus with a $\mathrm{^4He}$ gaseous target%
}

\author{~
\address{~}
\\
{A.~Scordo$^{1}$,C.~Amsler$^{2}$,M.~Bazzi$^{1}$,D.~Bosnar$^{3}$,M.~Bragadireanu$^{4}$,
M.~Cargnelli$^{2}$,M.~Carminati$^{5}$,A.~Clozza$^{1}$,G.~Deda$^{5}$,L.~De~Paolis$^{1}$,
R.~Del~Grande$^{6,1}$,L.~Fabbietti$^{6}$,C.~Fiorini$^{5}$,C.~Guaraldo$^{1}$,M.~Iliescu$^{1}$,
M.~Iwasaki$^{7}$,A.~Khreptak$^{1}$,P.~King$^{5}$,P.~Levi~Sandri$^{1}$,S.~Manti$^{1}$,
J.~Marton$^{5}$,M.~Miliucci$^{1}$,P.~Moskal$^{8}$,F.~Napolitano$^{1}$,S.~Nied\'zwiecki$^{8}$,
H.~Ohnishi$^{9}$,K.~Piscicchia$^{10}$,Y.~Sada$^{9}$,F.~Sgaramella$^{1\dag}$,H.~Shi$^{2}$,
M.~Silarski$^{8}$,D.L.~Sirghi$^{1,10,4}$,F.~Sirghi$^{1,4}$,M.~Skurzok$^{8,1}$,A.~Spallone$^{1}$,
K.~Toho$^{9}$,M.~T\"uchler$^{2}$,O.~Vazquez~Doce$^{1}$,C.~Yoshida$^{9}$,J.~Zmeskal$^{2}$
~and~C.~Curceanu$^{1}$
}
\address{
$^{\dag}$Corresponding author (email: Francesco.Sgaramella@lnf.infn.it),\\\vspace{1cm}
$^{1}$INFN Laboratori Nazionali di Frascati, Frascati (Roma), Italy\\
$^{2}$Stefan-Meyer-Institut f\"ur subatomare Physik, Vienna, Austria\\
$^{3}$Department of Physics, Faculty of Science, University of Zagreb, Croatia\\
$^{4}$Horia Hulubei National Institute of Physics and Nuclear Engineering (IFIN-HH), Magurele, Romania\\
$^{5}$Politecnico di Milano, Dipartimento di Elettronica, Informazione e Bioingegneria and INFN Sezione di Milano, Italy\\
$^{6}$Physik Department E62, Technische Universit\"at M\"unchen, Garching, Germany\\
$^{7}$RIKEN, Institute of Physical and Chemical Research, Wako, Japan\\
$^{8}$The M. Smoluchowski Institute of Physics, Jagiellonian University, 30-348 Krak\'ow, Poland\\
$^{9}$Research Center for Electron Photon Science (ELPH), Tohoku University, Sendai, Japan\\
$^{10}$Centro Ricerche Enrico Fermi-Museo Storico della fisica e Centro Studi e Ricerche “Enrico Fermi”, 00184 Roma, Italy.}
}

\maketitle
\begin{abstract}

\noindent In this paper, we present the first tests performed after the full installation of the SIDDHARTA-2 experimental apparatus on the Interaction Region of the 
DA$\mathrm{\Phi}$NE collider at the INFN National Laboratories of Frascati.
Before starting the first measurement of the kaonic deuterium $\mathrm{2p\rightarrow1s}$ transition,
accurate evaluation of the background rejection. mainly achieved with the Kaon Trigger system, was required.
This run, performed in the period 04-26/05/2022 with a $\mathrm{^4He}$ gaseous target, confirmed the $\mathrm{10^5}$ rejection factor obtained with a reduced version of the setup and different machine
conditions in 2021. This important outcome motivated the filling of the target cell with deuterium and the starting of the measurement campaign of the 
kaonic deuterium $\mathrm{2p\rightarrow1s}$ transition.
\end{abstract}
  
\section{Introduction}

\noindent X-ray spectroscopy of kaonic atoms is the perfect tool for the investigation
of the strong interaction in the low-energy limit \cite{Ramos:1999ku,Friedman:2012qy,Gal:2013vx,Ikeda:2012au,Gal:2016boi,Cieply:2011nq}.
In particular, an extraction of the $\mathrm{K^-p}$ and $\mathrm{K^-d}$ scattering lengths
with isospin-breaking corrections \cite{Doring:2011xc,Meissner:2006gx} is possible, through the Deser-Truemann-type formulae, 
by combining the measurements of the strong interaction induced
shifts and widths of the 1s level both in kaonic deuterium and hydrogen.\\
With this aim, soon after its successful measurement of kaonic hydrogen in 2009 \cite{SIDDHARTA:2011dsy}
at the DA$\mathrm{\Phi}$NE \cite{Zobov:2010zza} collider of the 
INFN Laboratories of Frascati, the SIDDHARTA collaboration proposed to realize an updated version of the experimental apparatus which, 
in April 2019, was ready to be installed on the Interaction Region (IR).\\
To perform both conditioning of the machine and tuning of the various components of the SIDDHARTA-2 setup, 
a reduced version, named SIDDHARTINO \cite{Sirghi:2022tvv}, with only 1/6 of the X-ray Silicon Drift Detectors (SDD) has been installed in 2019.
The SIDDHARTINO run started, due to the pandemic situation, only in January 2021 and two runs, with the target cell filled with $\mathrm{^4He}$ 
gas at about 1.5\% and 0.8\% of the liquid helium density, were performed to optimize the various components of the setup
as well as to provide feedback to the machine during its commissioning phase.
The choice of $\mathrm{^4He}$ was dictated by the high yield of the 
$\mathrm{K^4He(3d\rightarrow2p)}$ transition allowing for a very fast tuning. The experimental outcomes of this run already 
represented the first important physics results of the SIDDHARTA-2 experiment, delivering the most precise measurement of
the 2p level shift and width in a gaseous target \cite{Sirghi:2022tvv}. \\
In the second half of 2021 the full SIDDHARTA-2 setup was installed on the DA$\mathrm{\Phi}$NE IR \cite{Napolitano:2022eik}.
The increased number of SDDs, as well as the
different conditions of the machine background resulting from the optimization of the instantaneous luminosity, suggested performing 
a second test with helium before filling the target cell with the deuterium. This was indeed necessary to crosscheck the performances
of the experimental apparatus in its full version, with a particular focus on the background rejection capabilities of the trigger system.

\section{The SIDDHARTA-2 setup}\label{sec_setup}

\begin{figure}[htb]
\centerline{%
\includegraphics[width=11 cm]{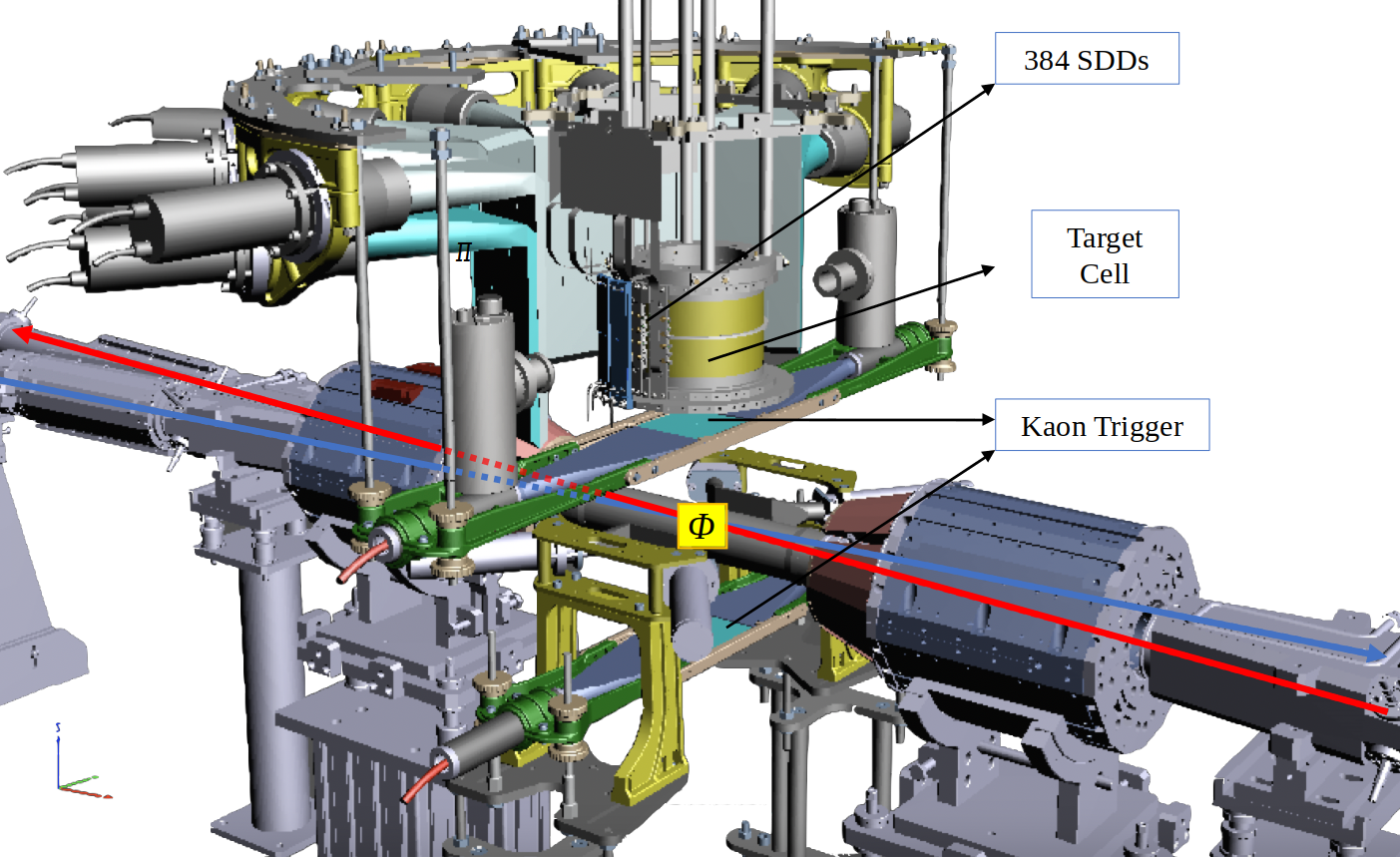}}
\caption{The SIDDHARTA-2 setup installed on the DA$\mathrm{\Phi}$NE Interaction Region at the INFN National Laboratories of Frascati.}
\label{Fig:setup}
\end{figure}

\noindent DA$\mathrm{\Phi}$NE is an $\mathrm{e^+e^-}$ collider with $\mathrm{e^+}$ and $\mathrm{e^-}$ 
beams tuned at $\mathrm{510\,MeV/c}$ momentum; this facility is then
a world-class $\mathrm{\Phi}$-factory, delivering low
energetic and monochromatic back-to-back $\mathrm{K^+K^-}$ pairs 
(16 MeV of kinetic energy) through the $\mathrm{\Phi}$-meson decay.
These properties render DA$\mathrm{\Phi}$NE the most suitable facility in the world
to perform high precision spectroscopy of kaonic atoms, 
as already evidenced by the many important results achieved in 2009 by 
the SIDDHARTA experiment \cite{SIDDHARTA:2011dsy,SIDDHARTA:2013ftj,SIDDHARTA:2009qht,SIDDHARTA:2010uae,SIDDHARTA:2012rsv}.\\
In Fig. \ref{Fig:setup} a drawing of the SIDDHARTA-2 apparatus is shown, where
the main components are highlighted. For this work, of particular interest are the Target Cell, the SDDs and the Kaon Trigger,
described below.

\noindent \emph{Target Cell (TC).} Enclosed in a cylindrical vacuum chamber and made of a high purity aluminum 
structure and a 150 µm thick Mylar wall, the TC is used to store the various targets. 
In particular, kaonic deuterium gas is planned to be stored at an equivalent density 
of 3\% of Liquid Deuterium (LD).\\ 
\emph{X-ray Silicon Drift Detectors (SDDs).} The X-rays emitted from the various transitions of kaonic atoms are measured by 
384 fast 450 $\mathrm{\mu m}$ thick SDDs, $\mathrm{0.64\,cm^2}$ surface each, arranged in 48 arrays (2 x 4 matrix), 
placed all around the target cell, for a total surface of $\mathrm{245.76\,cm^2}$.
Each unit is closely bonded to a C-MOS charge-sensitive amplifier (CUBE \cite{5873732}) and the signals are processed by a dedicated 
ASIC (SFERA, \cite{Quaglia:2016uox,7466864}). The SDDs system has been optimized in the laboratory and successfully
tested in the heavy background of the collider \cite{Sgaramella:2022rbl,Miliucci:2022lvn,Miliucci:2021vil,Miliucci:2021wbj}.
The SDDs are, together with the target cell, placed inside the vacuum chamber where they are cooled down to $\mathrm{-145^{\circ}C}$.\\
\emph{Kaon Trigger (KT).} This tool is used to identify those events
in which a kaon pair is delivered in the vertical direction with a characteristic Time of Flight (ToF), ensuring that a charged kaon entered in
the target cell. A dramatic reduction of the asynchronous component of the DA$\mathrm{\Phi}$NE machine background can be then achieved with the ToF selection.\\ 
Among the other components, a Luminosity Monitor is used to assess the luminosity delivered by the collider \cite{Skurzok:2020phi} while two VETO 
systems are implemented to reduce the hadronic background \cite{Bazzi:2013kwa,Tuchler:2018vyt}.\\
The Kaon Trigger is used to achieve a factor $\mathrm{10^5}$, without which signals from kaonic atoms would be impossible to be measured.\\
This work is focused on the tests performed immediately after the installation of the full apparatus on DA$\mathrm{\Phi}$NE with a
$\mathrm{^4He}$ target. 
A confirmation of the good performances achieved in the SIDDHARTINO run is then mandatory to declare the full SIDDHARTA-2 setup
ready for the kaonic deuterium measurement.

\section{The kaonic helium test}\label{sec_test}

\noindent The data presented in this paper have been collected in the period 04-26/05/2022 and correspond to $\mathrm{28\,pb^{-1}}$ of integrated luminosity. 
The SDDs were cooled down to $\mathrm{-145^{\circ}C}$, and the density of the target corresponded to about 1.4\% times the Liquid Helium one.
All the spectra are already calibrated in energy, with a procedure extensively discussed in \cite{Sgaramella:2022rbl}.
  
\subsection{Background suppression}\label{sec_sb}

\begin{figure}[htb]
\centering
\includegraphics[width=9.2 cm]{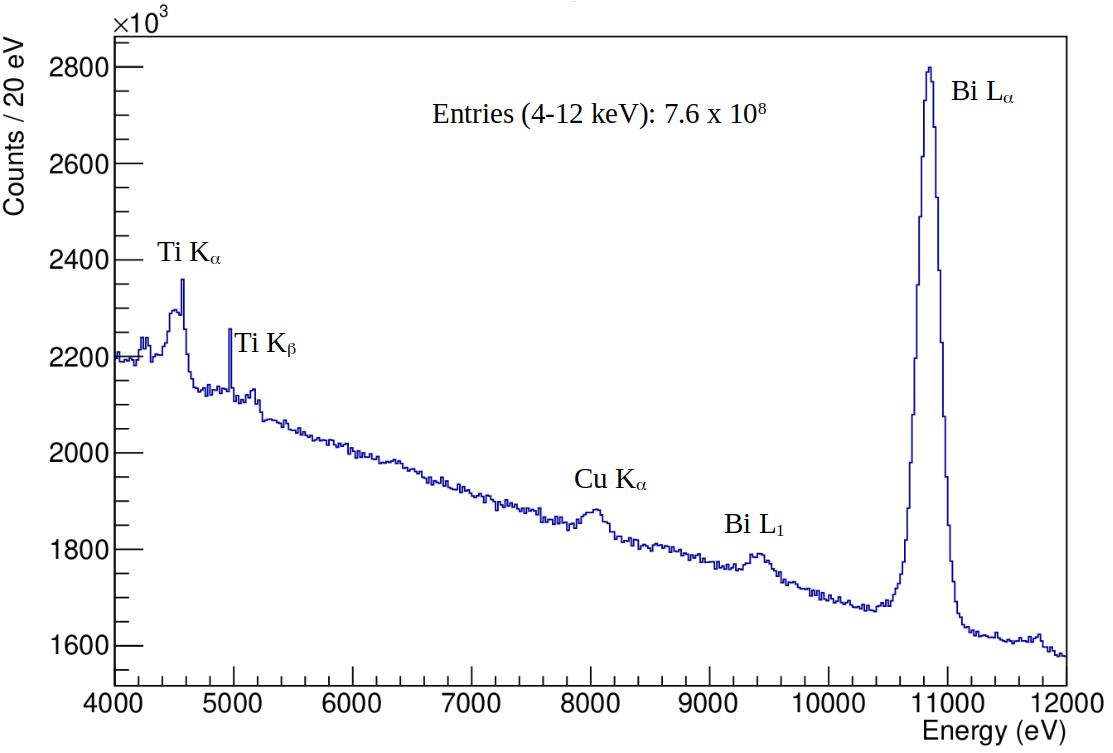}
\includegraphics[width=9.2 cm]{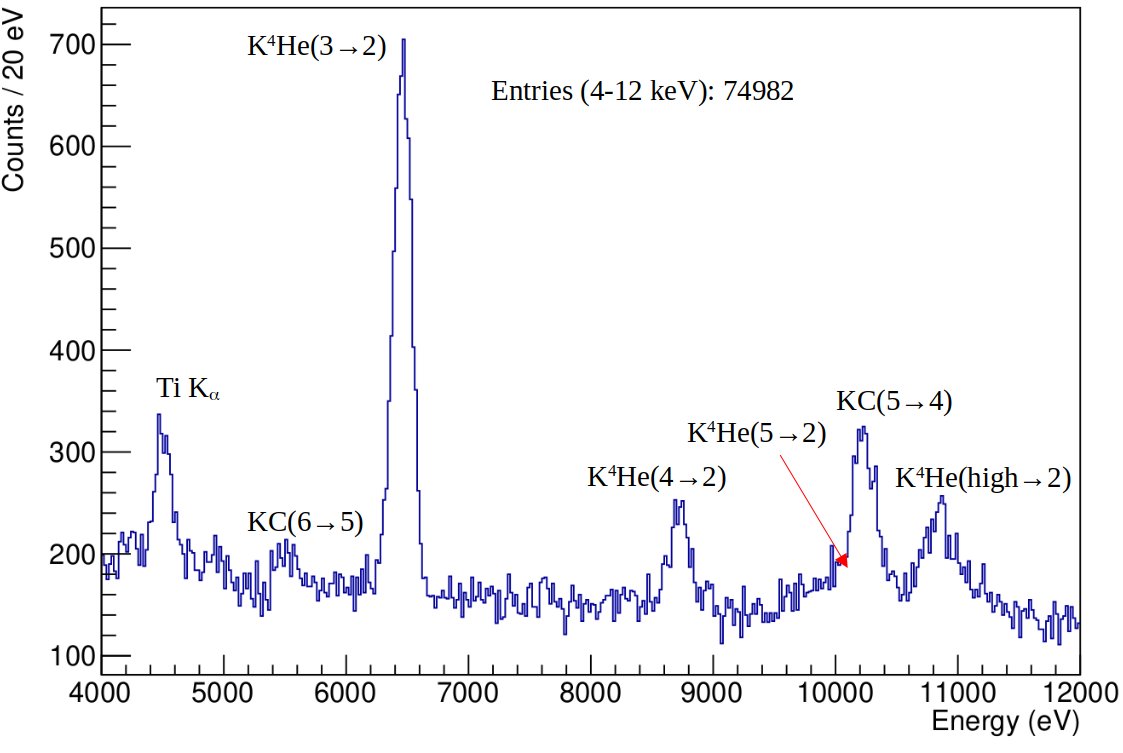}
\caption{\emph{Top}: The total $\mathrm{28\,pb^{-1}}$ spectrum obtained from the sum of the individual SDDs. 
Fluorescence peaks from Ti, Cu and Bi are present. \emph{Bottom:} Total spectrum after the trigger request (see text for details).}
\label{Fig:trigger_effect}
\end{figure}

\noindent In this section, the steps of the analysis from the raw data, collected by the SDDs, to the
final kaonic helium spectrum, are described.
The overall spectrum obtained without selection cut is shown in the upper pad of Fig. \ref{Fig:trigger_effect}, where peaks due to
the fluorescence of the various materials present in the experimental apparatus, thus not correlated in time with the beam crossing and the
$\mathrm{\Phi}$-decay, are present.\\
To remove this asynchronous background, a first selection is applied using the information from the KT; only triggered events in which 
two signals are detected in coincidence by the two scintillators are retained while all the others are discarded.\\
The effect of this selection is visible in the lower pad of Fig. \ref{Fig:trigger_effect}, where the number of events is drastically reduced and
the fluorescence peaks are not visible anymore, except for that from titanium which is present in the top of the target cell and is activated by
kaons not stopping in the gas. In the triggered spectrum the transitions of kaonic atoms formed in the
Mylar walls of the target are visible.

\begin{figure}[htb]
\centerline{%
\includegraphics[width=12.5 cm]{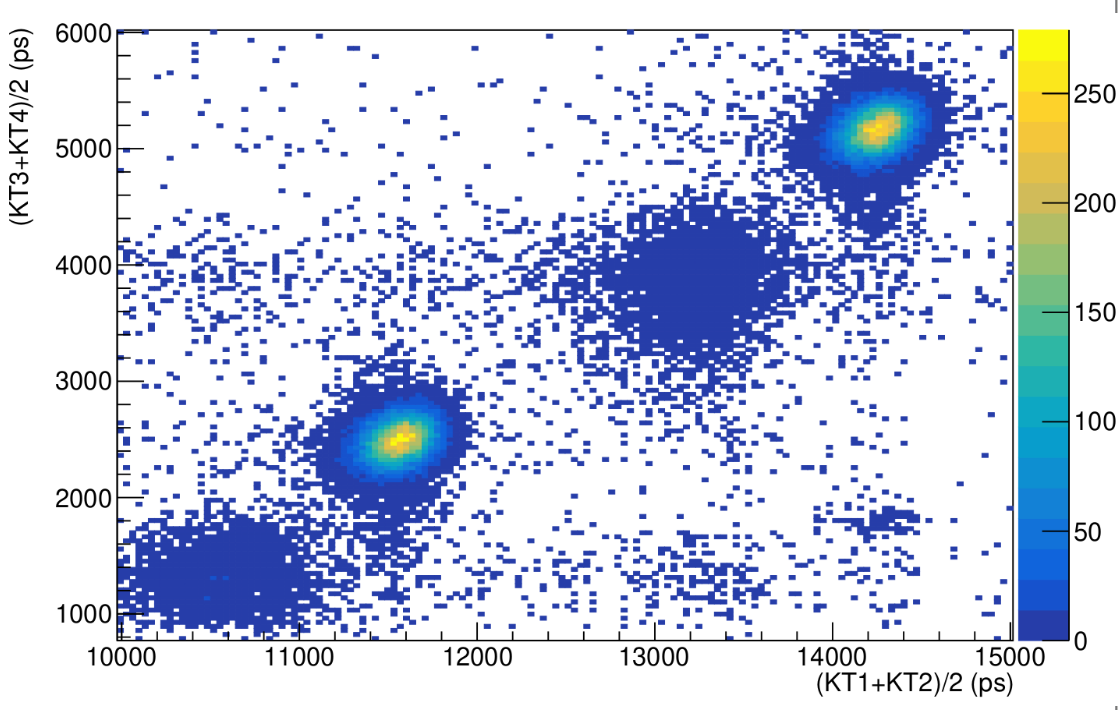}}
\caption{Scatter plot of the Mean Time of the two PMs reading the upper and lower scintillators of the Kaon Trigger. Clusters related to kaons and MIPs are
clearly visible in a double structure due to the usage of the RF/2 signal as time reference (see text for details).}
\label{Fig:kmip}
\end{figure}

\begin{figure}[htb]
\centerline{%
\includegraphics[width=12.5 cm]{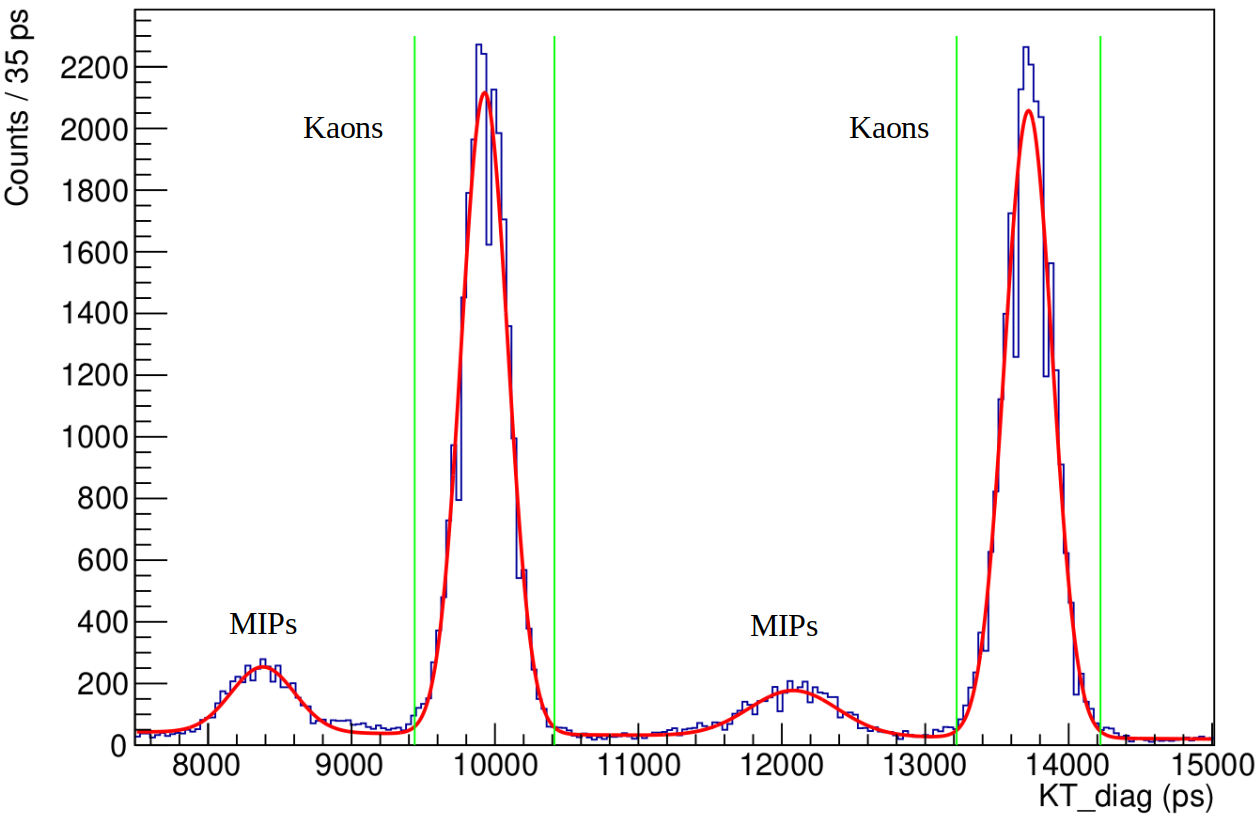}}
\caption{Projection on the diagonal of the scatter plot of Fig. \ref{Fig:kmip}. The fit (red line) is used to define the acceptance windows for the kaons,
while the double structure is due to the usage of the RF/2 as time reference (see text for details).}
\label{Fig:kmip_diag}
\end{figure}

\noindent The KT information could be also used to eliminate part of the remaining synchronous background, mainly consisting of Minimum Ionizing Particles (MIPs) produced 
in electromagnetic showers occurring within a fixed time window with the beam crossing.
The two scintillators of the KT allow discriminating by TOF between MIPs and kaons as shown in Fig. \ref{Fig:kmip} and \ref{Fig:kmip_diag}.
In Fig. \ref{Fig:kmip}, the scatter plot of the Mean Timers of the two PMs reading each scintillator is presented, where the distinct clusters corresponding to 
kaons (more intense) and MIPs (less intense) are visible in a double structure. 

\noindent The $\mathrm{\sim\,370\,MHz}$ radiofrequency (RF) of the $\mathrm{DA\Phi NE}$ collider, providing a trigger for every collision, 
and used as a time reference by the DAQ of the experiment, can't be handled by the most performing Constant Fracton Discriminators
(CFD), limited to work with 200 MHz, used to process it.
To overcome this limitation, the RF/2 is then used, at a frequency of $\mathrm{\sim\,185\,MHz}$; as a consequence, every coincidence event in the 
KT discriminators can be randomly associated in time with one of the two collisions.
The net results is the presence of the double structure visible both in Fig. \ref{Fig:kmip} and \ref{Fig:kmip_diag}.\\
In Fig. \ref{Fig:kmip_diag} the projection of the scatter plot on the diagonal is used to perform a fit and select only the events related to the kaons.

\begin{figure}[htb]
\centerline{%
\includegraphics[width=12.5 cm]{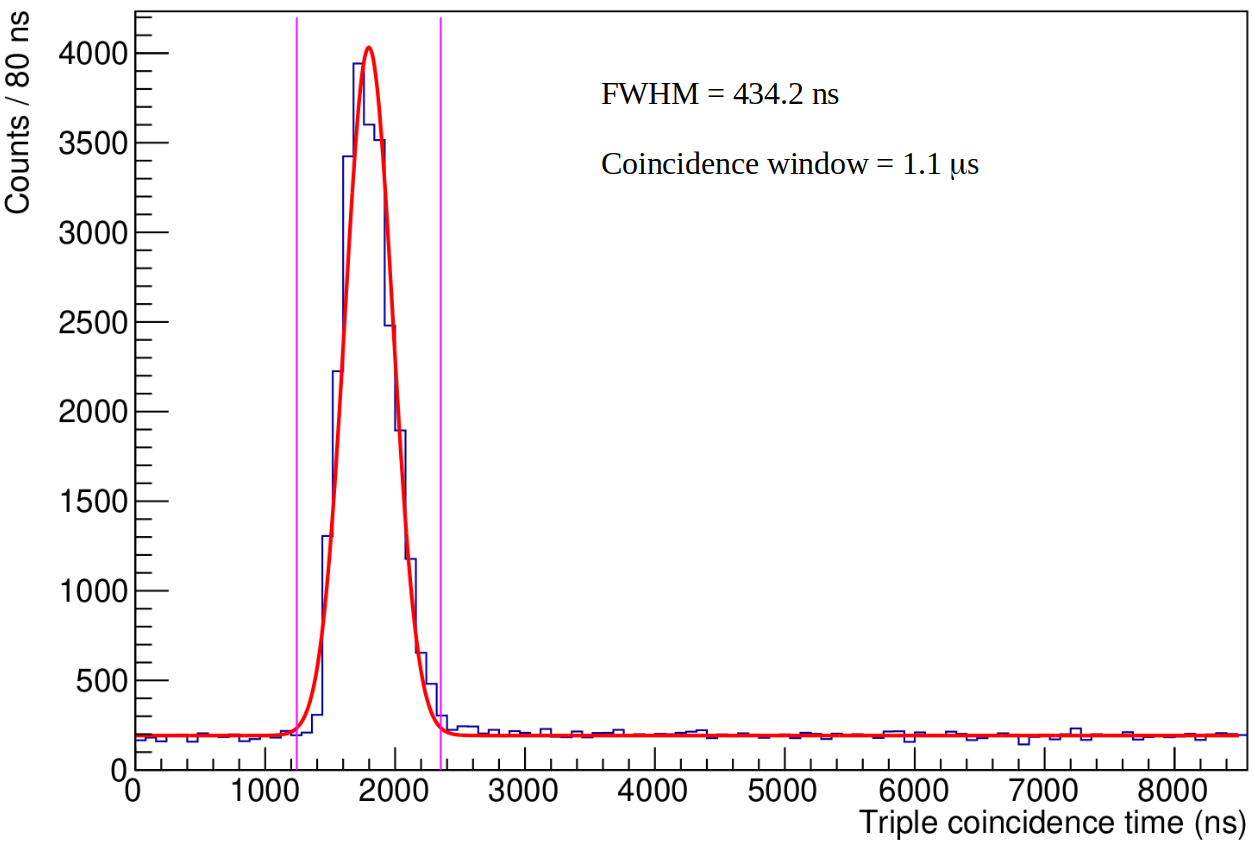}}
\caption{Time distribution of the triple coincidence between the two KT scintillators and a hit on an SDD. The peak width comes mainly from the $\mathrm{e^-}$ drift time
towards the anodes of the X-ray detectors (see text for details).}
\label{Fig:drift}
\end{figure}

\noindent As the last step, the time information from the SDDs can be exploited as well; a triple coincidence between the two scintillators of the KT and a hit 
in one of the SDDs, using the same RF/2 timestamp as reference time, is presented in Fig. \ref{Fig:drift}, where the peak represents the good signals while the flat 
background is due to accidental X-rays fake occurring randomly in time.\\
The drift time of the $\mathrm{e^-}$ towards the anode of an SDD after a photon detection, 
typically of the order of hundreds of ns, is much larger than the time needed by the kaons to reach the target, to be then moderated in the gas, to form the kaonic atom 
and finally go through the cascade emitting X-rays.
For this reason, the shape of the timing distribution in Fig. \ref{Fig:drift} is mainly due to the Drift Time of the $\mathrm{e^-}$ in the SDDs \cite{Miliucci:2022lvn}.
A further event selection can be performed by rejecting all the events lying outside the $\mathrm{1.1\,\mu s}$ window shown in Fig. \ref{Fig:drift}.
It has to be stressed that, with respect to the $\mathrm{-100^{\circ}C}$ of the SIDDHARTINO run, the $\mathrm{-145^{\circ}C}$ cooling of the SDDs allows obtaining a much narrower time window than the almost $\mathrm{2\,\mu s}$ previous one, thanks to a reduction of the FWHM of the triple coincidence peak from 950 ns to about 430 ns.

\begin{figure}[htb]
\centerline{%
\includegraphics[width=12.5 cm]{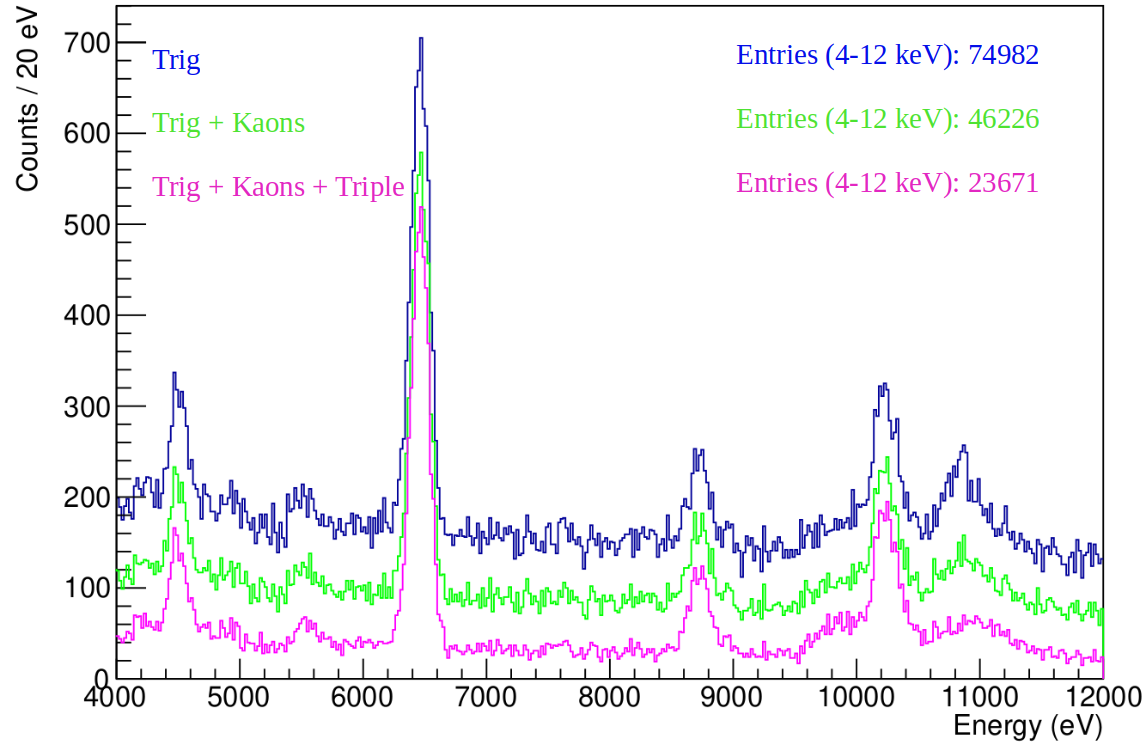}}
\caption{Overimposed total $\mathrm{28\,pb^{-1}}$ spectra obtained from the sum of the individual SDDs after the trigger requirement, 
the kaon selection and the application of triple coincidence window.}
\label{Fig:drift_eff}
\end{figure}

\noindent The overall effects of all these selections are presented in Fig. \ref{Fig:drift_eff} where the spectra with only the KT flag, the further addition of the 
kaon selection and then the triple coincidence requirement are shown in black, green and magenta, respectively.
The number of events for each spectrum are reported; comparing these values with those presented in Fig. \ref{Fig:trigger_effect}, 
a background rejection factor of $\mathrm{3.1\times10^-5}$ can be extracted. 
The results confirm those obtained during the SIDDHARTINO run \cite{Sirghi:2022tvv}.

\section{Conclusions and perspectives}\label{sec_conclusions}

\noindent In this paper, we presented the first tests performed after the full installation of the SIDDHARTA-2 experimental apparatus on the Interaction Region of the 
DA$\mathrm{\Phi}$NE collider at the INFN National Laboratories of Frascati.
The test, performed in the period 04-26/05/2022, has been used to confirm the very good performances of the trigger system achieved during the SIDDHARTINO test run,
performed in 2021 with a reduced number of X-ray detectors and different machine conditions.
This test was mandatory to assess that these new conditions, as well as the larger amount of detectors, were not negatively impacting on the crucial background rejection factor.
The outcomes of this run confirmed the $\mathrm{10^5}$ rejection factor obtained with SIDDHARTINO \cite{Sirghi:2022tvv} and allows to start the first 
measurement of the kaonic deuterium $\mathrm{2p\rightarrow1s}$ transition, which will be performed in 2022 and 2023.

\section*{Acknowledgments}

\noindent We thank C. Capoccia from LNF-INFN and H. Schneider, L. Stohwasser, and D. Pristauz- Telsnigg from Stefan-
Meyer-Institut for their fundamental contribution in designing and building the SIDDHARTA-2 setup. We thank as
well the $\mathrm{DA\Phi NE}$ staff for the excellent working conditions and permanent support. Part of this work was supported by
the Austrian Science Fund (FWF): [P24756-N20 and P33037-N]; the Croatian Science Foundation under the project
IP-2018-01-8570; the EU STRONG-2020 project (Grant Agreement No. 824093), the EU Horizon 2020 project
under the MSCA (Grant Agreement 754496); Japan Society for the Promotion of Science JSPS KAKENHI Grant
No. JP18H05402; the Polish Ministry of Science and Higher Education grant No. 7150/E-338/M/2018 and the Polish
National Agency for Academic Exchange (grant no PPN/BIT/2021/1/00037).

\bibliographystyle{Science}
\bibliography{jagsymp2022}
\end{document}